# Stationary bound states of spin-half particles in the Kerr and Kerr-Newman gravitational fields

M.V. Gorbatenko, V.P. Neznamov[1]

RFNC-VNIIEF, 37 Mira Ave., Sarov, 607188, Russia

Abstract

We prove the possibility of existence of stationary bound states of spin-half particles for the Kerr and Kerr-Newman gravitational fields using Chandrasekhar's Hamiltonian.

If the Hilbert condition is satisfied, $g_{00} > 0$, bound states of Dirac particles with a real discrete energy spectrum can exist both for particles outside the surface of the outer ergosphere of the Kerr and Kerr-Newman fields, and for particles under the surface of the inner ergosphere.

In this case, the surfaces of the external and internal ergospheres play the role of infinitely high potential barriers. Spin-half quantum-mechanical particles cannot cross the ergosphere surfaces of the Kerr and Kerr-Newman fields.

Based on the results of this study, we can assume that there exists a new type of rotating collapsars, for which the Hawking radiation is absent.

The results of this study can lead to a revision of some concepts of the standard cosmological model related to the evolution of the universe and interaction of rotating collapsars with surrounding matter.

---

[1] E-mail: neznamov@vniief.ru

# 1. Introduction

In [1] - [3], we developed an algorithm for deriving self-conjugate Dirac Hamiltonians with a flat scalar product of wave functions within the framework of pseudo-Hermitian quantum mechanics for arbitrary, including time dependent, external gravitational fields.

It follows from single-particle quantum mechanics that if the Hamiltonian is Hermitian with the appropriate equality of scalar products of the wave functions $\left((\Phi, H\Psi) = (H\Phi, \Psi)\right)$ and if appropriate boundary conditions are specified, the time-independent self-conjugate Hamiltonians $\left(H = H^+\right)$ should provide for the existence of stationary bound states of spin-half particles with a real energy spectrum.

In [4], [5], in numerical calculations with the use of the self-conjugate Hamiltonian with the flat scalar product of the wave functions, we for the first time obtained a infinite set of stationary non-decaying bound states of spin-half particles in the Schwarzschild gravitational field for any values of the gravitational coupling constant.

In order to meet the Hilbert condition, $g_{00} > 0$, we introduced a boundary condition such that components of the vector of current density of Dirac particles are zero near the "event horizon".

In [6], a similar approach was applied to the quantum-mechanical behavior of Dirac particles in the Reissner-Nordström gravitational field. As a result of the analysis, we concluded that bound states of spin-half particles with a real discrete energy spectrum can exist both above the external "event horizon" and under the internal "event horizon", the Cauchy horizon.

If the condition $g_{00} > 0$ is fulfilled, the "event horizons" of the Schwarzschild and Reissner-Nordström fields in the quantum-mechanical framework represent infinitely high potential barriers that deprive probe Dirac particles of the opportunity to cross them. It follows from this that such collapsars cannot radiate by the Hawking mechanism [7].

In this study, similarly to [4] - [6], we explore the possibility of existence of stationary bound states of Dirac particles in the Kerr [8] and Kerr-Newman [9] gravitational fields. As a result of the analysis, we conclude that bound states of spin-half particles with a real energy spectrum can exist both above the surfaces of the external ergospheres of the Kerr and Kerr-Newman fields and under the surfaces of the internal ergospheres.

Results of numerical calculations of the energy spectrum and wave functions will be presented in our next papers.

The present paper has the following structure. Sect. 2 is devoted to the quantum-mechanical behavior of Dirac particles in the Kerr field. In Sects. 2.1, 2.2, we introduce the notation and give the Kerr solution in Boyer-Lindquist coordinates and two Dirac Hamiltonians. One of them is a self-conjugate Hamiltonian with a flat scalar product of wave functions derived by us in [3]. The second is physically equivalent Chandrasekhar's Hamiltonian derived in [10].

As far as Chandrasekhar in [10] separated angular and radial variables in the Dirac equation for the Kerr field, in Sect. 2.3 we analyze the equations and asymptotics for radial wave functions.

In Sect. 2.4, we discuss the issues of the Hamiltonian's Hermiticity, boundary conditions for the wave functions and conclude that stationary bound states of Dirac particles with a real energy spectrum exist.

In Sect. 2.5, we discuss the cases of an extreme Kerr field and a naked singularity.

In Sect. 3, we in a similar manner explore the possibility of existence of stationary bound states of Dirac particles in the Kerr-Newman field.

In Conclusions we summarize the outcome of our analysis.

## 2. Kerr gravitational field

### 2.1. Kerr metric in the Boyer-Lindquist coordinates

Kerr's solution of the general relativity equations is characterized by a point source of gravitational field of mass $M$, which rotates with an angular momentum of $\mathbf{J} = Mc\mathbf{a}$, where $c$ is the speed of light.

Below we use the system of units $\hbar = c = 1$.

Tetrads are defined by the relationship

$$H_{\underline{\alpha}}^{\mu} H_{\underline{\beta}}^{\nu} g_{\mu\nu} = \eta_{\underline{\alpha\beta}}, \tag{1}$$

where

$$\eta_{\underline{\alpha\beta}} = diag[1, -1, -1, -1]. \tag{2}$$

Global indices (non-underlined) are lowered and raised up by means of the metric tensor $g_{\mu\nu}$ and inverse tensor $g^{\mu\nu}$, and the local (underlined) indices, by means of the tensors $\eta_{\underline{\alpha\beta}}$, $\eta^{\underline{\alpha\beta}}$.

Dirac matrices satisfy the relationships

$$\gamma^{\mu}\gamma^{\alpha} + \gamma^{\alpha}\gamma^{\mu} = 2g^{\mu\alpha}E, \tag{3}$$

$$\gamma^{\underline{\mu}}\gamma^{\underline{\alpha}} + \gamma^{\underline{\alpha}}\gamma^{\underline{\mu}} = 2\eta^{\underline{\mu\alpha}}E, \tag{4}$$

where $E$ is a 4 x 4 unity matrix.

The relationship between $\gamma^\alpha$ and $\gamma^{\underline{\alpha}}$ is given by the expression

$$\gamma^\alpha = H^\alpha_{\underline{\beta}} \gamma^{\underline{\beta}}. \tag{5}$$

The Kerr solution in the Boyer-Lindquist coordinates $(t, r, \theta, \varphi)$ is given by

$$ds^2 = \left(1 - \frac{r_0 r}{\rho_K^2}\right) dt^2 + \frac{2 a r_0 r}{\rho_K^2} \sin^2\theta \, dt d\varphi - \frac{\rho_K^2}{\Delta} dr^2 - \rho_K^2 d\theta^2 - \\ - \left(r^2 + a^2 + \frac{a^2 r_0 r}{\rho_K^2} \sin^2\theta\right) \sin^2\theta \, d\varphi^2. \tag{6}$$

In (6), $r_0 = \frac{2GM}{c^2}$ is the gravitational radius ("event horizon") of the Schwarzschild field, $G$ is the gravitational constant, $\rho_K^2 = r^2 + a^2 \cos^2\theta$, $\Delta = r^2 - r_0 r + a^2$.

In accordance with the Hilbert condition $(g_{00} > 0)$ in (6) it is meant that the following inequality holds:

$$\left(1 - \frac{r_0 r}{\rho_K^2}\right) > 0. \tag{7}$$

In the $(r, \theta)$ coordinates, the equality of (7) to zero determines the outer and the inner surfaces of ergospheres of the Kerr field.

The inverse tensor $g^{\alpha\beta}$ has the following form:

$$g^{\alpha\beta} = \begin{array}{|c|c|c|c|} \hline \frac{1}{\Delta}\left(r^2 + a^2 + \frac{a^2 r_0 r}{\rho_K^2}\sin^2\theta\right) & 0 & 0 & \frac{a r_0 r}{\Delta \rho_K^2} \\ \hline 0 & -\frac{\Delta}{\rho_K^2} & 0 & 0 \\ \hline 0 & 0 & -\frac{1}{\rho_K^2} & 0 \\ \hline \frac{a r_0 r}{\Delta \rho_K^2} & 0 & 0 & -\frac{1}{\Delta \sin^2\theta}\left(1 - \frac{r_0 r}{\rho_K^2}\right) \\ \hline \end{array} \tag{8}$$

## 2.2. Hamiltonians of spin-half particles in the Kerr field

In [3], for the solution (6) we obtained a self-conjugate Hamiltonian in the $\eta$-representation with a flat scalar product of wave functions. It is written in a fairly complicated form:

$$H_\eta = \frac{m}{\sqrt{g^{00}}}\gamma^0 - \frac{i\sqrt{\Delta}}{\rho_K\sqrt{g^{00}}}\gamma^0\gamma^1\left(\frac{\partial}{\partial r}+\frac{1}{r}\right) - \frac{i}{\rho_K\sqrt{g^{00}}}\gamma^0\gamma^2\left(\frac{\partial}{\partial\theta}+\frac{1}{2}\mathrm{ctg}\,\theta\right) -$$
$$-\frac{i}{g^{00}\sqrt{\Delta}\sin\theta}\gamma^0\gamma^3\frac{\partial}{\partial\varphi} - \frac{i}{g^{00}}\frac{ar_0 r}{\rho_K^2\Delta}\frac{\partial}{\partial\varphi} - \frac{i}{2}\gamma^0\gamma^1\left[\frac{\partial}{\partial r}\frac{\sqrt{\Delta}}{\rho_K\sqrt{g^{00}}}\right] -$$
$$-\frac{i}{2}\gamma^0\gamma^2\left[\frac{\partial}{\partial\theta}\frac{1}{\rho_K\sqrt{g^{00}}}\right] + \frac{i}{4}\gamma^3\gamma^1\sqrt{g^{00}}\frac{\Delta}{\rho_K}ar_0\sin\theta\frac{\partial}{\partial r}\left(\frac{r}{g^{00}\rho_K^2\Delta}\right) -$$
$$-\frac{i}{4}\gamma^2\gamma^3\sqrt{g^{00}}\frac{\sqrt{\Delta}}{\rho_K}ar_0\sin\theta\frac{\partial}{\partial\theta}\left(\frac{r}{g^{00}\rho_K^2\Delta}\right). \tag{9}$$

If in (9) we restrict ourselves to linear terms with respect to $a$, we will obtain a self-conjugate Hamiltonian for a weak Kerr field.

$$H_\eta^{app} = m\sqrt{f_S}\gamma^0 - if_S\gamma^0\gamma^1\left(\frac{\partial}{\partial r}+\frac{1}{r}\right) - \frac{ir_0}{2r^2}\gamma^0\gamma^1 -$$
$$-i\sqrt{f_S}\gamma^0\left[\gamma^2\frac{1}{r}\left(\frac{\partial}{\partial\theta}+\frac{1}{2}\mathrm{ctg}\,\theta\right) + \gamma^3\frac{1}{r\sin\theta}\frac{\partial}{\partial\varphi}\right] - \frac{iar_0}{r^3}\frac{\partial}{\partial\varphi} - \tag{10}$$
$$-i\frac{3}{4}\frac{ar_0}{r^3}\sin\theta\gamma^3\gamma^1.$$

In (10) $f_S = 1 - \frac{r_0}{r}$.

With $a = 0$, the Hamiltonians (9), (10) coincide with the self-conjugate Hamiltonian in the Schwarzschild field [3] - [5].

As distinct from the centrally symmetric Schwarzschild, Reissner-Nordström and other gravitational fields, the axially symmetric Kerr field does not allow us to use spin-half spherical harmonics for separating angular variables in the Dirac equation.

Chandrasekhar in his paper [10] separated variables in the Dirac Hamiltonian with the metric (6) using the Penrose-Newman two-component spinor formalism [11] and the Kinnersley unit tetrad [12].

Following [13], [14], the Dirac equation and the Chandrasekhar Hamiltonian can be written in the bispinor form:

$$i\frac{\partial\psi_{Ch}}{\partial t} = H_{Ch}\psi_{Ch} = \left(\frac{m}{g^{00}}\gamma^0 - \frac{i}{g^{00}}\gamma^0\gamma^k\frac{\partial}{\partial x^k} - i\Phi^0 - \frac{i}{g^{00}}\gamma^0\gamma^k\Phi^k\right)\psi_{Ch}. \tag{11}$$

In (11), $k = 1, 2, 3$, $\Phi^0$, $\Phi^k$ are bispinor connectivities calculated in a standard way.

The remaining quantities have the following form:

$$\psi_{Ch} = \begin{pmatrix} P^A \\ Q_B^* \end{pmatrix}. \tag{12}$$

In (11),

$$\gamma^\mu = \begin{pmatrix} 0 & \sqrt{2}\sigma^{\mu AB'} \\ \sqrt{2}\left[\sigma^\mu_{AB'}\right]^T & 0 \end{pmatrix}, \qquad (13)$$

$$\sqrt{2}\left[\sigma^{\mu AB'}\right] = \sqrt{2}\begin{pmatrix} n^\mu & -m^{*\mu} \\ -m^\mu & l^\mu \end{pmatrix}; \quad \sqrt{2}\left[\sigma^\mu_{AB'}\right]^T = \sqrt{2}\begin{pmatrix} l^\mu & m^{*\mu} \\ m^\mu & n^\mu \end{pmatrix}. \qquad (14)$$

In (13), (14), the indices $A, B'$ assume the values of 0 and 1, and the signs $*$ and $T$ mean complex conjugation and transposition.

Components of the Kinnersley tetrad equal

$$l^\mu = \frac{1}{\sqrt{2}}\left(\frac{r^2+a^2}{\Delta}, 1, 0, \frac{a}{\Delta}\right), \quad n^\mu = \frac{1}{\sqrt{2}\rho_K^2}\left[r^2+a^2, -\Delta, 0, a\right]$$
$$m^\mu = \frac{1}{\sqrt{2}(r+ia\cos\theta)}\left(ia\sin\theta, 0, 1, \frac{i}{\sin\theta}\right). \qquad (15)$$

The inverse metric tensor (8) is expressed in terms of the components of (15) as follows:

$$g^{\mu\nu} = l^\mu n^\nu + n^\mu l^\nu - m^\mu m^{*\nu} - m^{*\mu} m^\nu. \qquad (16)$$

Chandrasekhar's Hamiltonian (11) is physically equivalent to the self-conjugate Hamiltonian (9), because they are related by a similarity transformation.

For the wave functions of the Dirac equation with Chandrasekhar's Hamiltonian (11), the scalar product contains Parker's weight operator [15], [1] - [3].

$$\rho_P = \sqrt{g_G}\gamma^0\gamma^0, \qquad (17)$$

where for the Kerr solution in the Boyer-Lindquist coordinates, $g_G = \dfrac{\rho_K^4}{r^4}$ [3].

If the self-conjugate Hamiltonian (9) is used, $\rho_P = 1$.

If we define the operator $\eta$ from the equality

$$\rho_P = \eta^+\eta, \qquad (18)$$

then Chandrasekhar's Hamiltonian will be related to the Hamiltonian (9) by the similarity transformation

$$H_\eta = \eta H_{Ch}\eta^{-1}. \qquad (19)$$

If follows from (19) that both Hamiltonians have the same energy spectrum.

In the general case, the expression for the operator $\eta$ is complex and cumbersome. If $a = 0$ (Schwarzschild field), the operator $\eta$ is diagonal and has the following form:

$$\eta = diag\left[\frac{1}{\sqrt{f_s}},1,1,\frac{1}{\sqrt{f_s}}\right]. \quad (20)$$

Considering the above, to analyze the possibility of existence of stationary bound states of spin-half particles in the Kerr field, below we use the procedure of separation of variables implemented by Chandrasekhar in [10] (see also [13]).

For the stationary case, the wave function in (11) can be written as

$$\psi(\mathbf{r},t) = \psi(\mathbf{r})e^{-iEt}, \quad (21)$$

where $E$ is the energy of the Dirac particle.

Then, representing the function (21) as

$$\psi(\mathbf{r},t) = \begin{pmatrix} \dfrac{1}{(r-ia\cos\theta)}\overset{(-)}{R}(r)\overset{(-)}{S}(\theta) \\ \dfrac{1}{\sqrt{\Delta}}\overset{(+)}{R}(r)\overset{(+)}{S}(\theta) \\ -\dfrac{1}{\sqrt{\Delta}}\overset{(+)}{R}(r)\overset{(-)}{S}(\theta) \\ -\dfrac{1}{(r+ia\cos\theta)}\overset{(-)}{R}(r)\overset{(+)}{S}(\theta) \end{pmatrix} e^{-iEt}e^{im_\varphi \varphi} \quad (22)$$

we can separately obtain two equations for the angular functions $\overset{(-)}{S}(\theta), \overset{(+)}{S}(\theta)$ and two equations for the radial functions $\overset{(-)}{R}(r), \overset{(+)}{R}(r)$:

$$\left(\frac{d}{d\theta} + aE\sin\theta - m_\varphi \frac{1}{\sin\theta} + \frac{1}{2}\operatorname{ctg}\theta\right)\overset{(-)}{S}(\theta) = (\lambda + am\cos\theta)\overset{(+)}{S}(\theta),$$
$$\left(\frac{d}{d\theta} - aE\sin\theta + m_\varphi \frac{1}{\sin\theta} + \frac{1}{2}\operatorname{ctg}\theta\right)\overset{(+)}{S}(\theta) = (-\lambda + am\cos\theta)\overset{(-)}{S}(\theta), \quad (23)$$

$$\Delta^{\frac{1}{2}}\left(\frac{d}{dr} - i\frac{K}{\Delta}\right)\overset{(-)}{R}(r) = (\lambda + imr)\overset{(+)}{R}(r),$$
$$\Delta^{\frac{1}{2}}\left(\frac{d}{dr} + i\frac{K}{\Delta}\right)\overset{(+)}{R}(r) = (\lambda - imr)\overset{(-)}{R}(r). \quad (24)$$

In Eqs. (24),

$$K = (r^2 + a^2)E - m_\varphi a. \quad (25)$$

In Eqs. (23), (24), $\lambda$ is the separation constant. As distinct from the centrally symmetric Schwarzschild and Reissner-Nordström fields, in addition to the standard dependence of $\lambda$ on the quantum numbers $j,l$, in the Kerr field, $\lambda$ also depends on the magnetic quantum number $m_\varphi$, on the angular momentum $a$, on energy and mass of the probe particle. Thus, when solving the equations (24), for each value of energy $E$, momentum $a$, quantum number $m_\varphi$ in order to determine the parameter $\lambda$, Eqs. (23) also must be solved with appropriate boundary conditions.

In his dissertation [14], Dolan numerically determined the separation constant $\lambda$ in the range of the values of $aE$ from zero to eight and up to $j = \dfrac{11}{2}$.

The angular functions $\overset{(-)}{S}(\theta), \overset{(+)}{S}(\theta)$ are spin-weighted spheroidal harmonics. An overview of their wide application in theoretical physics can be found, for example, in [14].

The spheroidal harmonics satisfy the following symmetry relationships:

$$\overset{(-)}{S}(\theta) = \overset{(+)}{S}(\pi - \theta)$$
$$S_{l,m_\varphi,E}^{j=l+\frac{1}{2}}(\theta) = (-1)^{l+m_\varphi} S_{l,m_\varphi,E}^{j=l-\frac{1}{2}}(\pi - \theta). \tag{26}$$

$$S_{j,l,m_\varphi,E}(\theta) = (-1)^j S_{j,l,-m_\varphi,-E}(\pi - \theta), \tag{27}$$

$$S_{l,m_\varphi,E}^{j=l+\frac{1}{2}}(\theta) = (-1)^{m+\frac{1}{2}} S_{l,-m_\varphi,-E}^{j=l-\frac{1}{2}}(\theta). \tag{28}$$

## 2.3. Equations and asymptotics for radial wave functions

As far as it follows from the system of equations (24) that $\overset{(+)}{R}{}^*(r) = \overset{(-)}{R}(r)$, we obtain equations for real radial functions

$$g(r) = \overset{(-)}{R}(r) + \overset{(+)}{R}(r)$$
$$f(r) = -i\left(\overset{(-)}{R}(r) - \overset{(+)}{R}(r)\right). \tag{29}$$

Equations for the functions $f(r)$ and $g(r)$ have the following form:

$$f_K \frac{d}{dr} f + \frac{\sqrt{f_K}}{r} \lambda f - \left(E\left(1 + \frac{a^2}{r^2}\right) - \frac{m_\varphi a}{r^2} + \sqrt{f_K}\, m\right) g = 0$$
$$f_K \frac{d}{dr} g - \frac{\sqrt{f_K}}{r} \lambda g + \left(E\left(1 + \frac{a^2}{r^2}\right) - \frac{m_\varphi a}{r^2} - \sqrt{f_K}\, m\right) f = 0. \tag{30}$$

In Eqs. (30),

$$f_K = 1 - \frac{r_0}{r} + \frac{a^2}{r^2}. \tag{31}$$

The condition (7) leads to the necessity of considering only positive values of $f_K > 0$. The quantity $f_K$ in (31) can be represented as

$$f_K = \left(1 - \frac{r_+}{r}\right)\left(1 - \frac{r_-}{r}\right), \qquad (32)$$

where

$$r_\pm = \frac{r_0}{2} \pm \sqrt{\frac{r_0^2}{4} - a^2}. \qquad (33)$$

With $r_0^2 \geq 4a^2$, the quantities $r_\pm$ are outer and inner radii of the "event horizons" for the Kerr field.

We introduce dimensionless variables $\varepsilon = \frac{E}{m}$; $\rho = \frac{r}{l_c}$; $2\alpha = \frac{r_0}{l_c} = \frac{2GMm}{\hbar c}$; $\alpha_a = \frac{a}{l_c}$; where $l_c = \frac{\hbar}{mc}$ is the Compton wavelength of the Dirac particle.

In the dimensionless variables, the quantities (32), (33) can be represented as

$$f_K = \left(1 - \frac{\rho_+}{\rho}\right)\left(1 - \frac{\rho_-}{\rho}\right), \qquad (34)$$

$$\rho_+ = \alpha + \sqrt{\alpha^2 - \alpha_a^2}, \qquad (35)$$

$$\rho_- = \alpha - \sqrt{\alpha^2 - \alpha_a^2}. \qquad (36)$$

Eqs. (30) in the dimensionless variables have the following form:

$$\left(1 - \frac{\rho_+}{\rho}\right)\left(1 - \frac{\rho_-}{\rho}\right)\frac{df}{d\rho} + \lambda \frac{\sqrt{\left(1 - \frac{\rho_+}{\rho}\right)\left(1 - \frac{\rho_-}{\rho}\right)}}{\rho} f - \left(\varepsilon\left(1 + \frac{\alpha_a^2}{\rho^2}\right) - \frac{m_\varphi \alpha_a}{\rho^2} + \sqrt{\left(1 - \frac{\rho_+}{\rho}\right)\left(1 - \frac{\rho_-}{\rho}\right)}\right) g = 0$$

$$\left(1 - \frac{\rho_+}{\rho}\right)\left(1 - \frac{\rho_-}{\rho}\right)\frac{dg}{d\rho} - \lambda \frac{\sqrt{\left(1 - \frac{\rho_+}{\rho}\right)\left(1 - \frac{\rho_-}{\rho}\right)}}{\rho} g + \left(\varepsilon\left(1 + \frac{\alpha_a^2}{\rho^2}\right) - \frac{m_\varphi \alpha_a}{\rho^2} - \sqrt{\left(1 - \frac{\rho_+}{\rho}\right)\left(1 - \frac{\rho_-}{\rho}\right)}\right) f = 0$$

(37)

As far as the values of $f_K > 0$ in (30), (37) are possible only if $\rho > \rho_+$ and $\rho < \rho_-$, the wave functions in the range between the outer and inner "event horizons" $(\rho_- \leq \rho \leq \rho_+)$ are zero.

At $\rho > \rho_+$, following the condition (7), the domain for the wave functions of the Dirac equation in the Kerr field is the space $(\rho, \theta, \varphi)$ with the radii $\rho(\theta)$ larger than the radii of the surface of the outer ergosphere. If $\rho < \rho_-$, the domain for the wave functions, on the contrary, is the range with the radii $\rho(\theta)$ smaller than the radii of the surface of the inner ergosphere.

If the angular momentum $a$ is zero, i.e. $\alpha_a = 0$, then $\rho_+ = 2\alpha$; $\rho_- = 0$. In this case, Eqs. (30), (37) coincide with the system of radial equations for the Schwarzschild field with one "event horizon" ($r = r_0$ or $\rho = 2\alpha$); the separation constant $\lambda$ does not depend on the particle energy and becomes equal to the separation constant $(-\kappa)$ in the system of Dirac equations in the external field of the Coulomb potential

$$\kappa = \pm 1, \pm 2 ... = \begin{cases} -(l+1), & j = l + \frac{1}{2} \\ l, & j = l - \frac{1}{2} \end{cases}. \tag{38}$$

Consider the asymptotics of the radial wave functions $f(\rho), g(\rho)$ with $\rho \to \infty$; $\rho \to \rho_+ (\rho > \rho_+)$; $\rho \to \rho_- (\rho < \rho_-)$; $\rho \to 0$.

With $\rho \to \infty$, the asymptotic behavior of the wave functions $f(\rho), g(\rho)$ for finite motion is the same as for the centrally symmetric Schwarzschild, Reissner-Nordström and other gravitational fields [4], [5], [6].

For $\rho \to \infty$ leading terms of asymptotics equal

$$\begin{aligned} f &\to C e^{-\rho\sqrt{1-\varepsilon^2}} \\ g &\to -\sqrt{\frac{1-\varepsilon}{1+\varepsilon}} f. \end{aligned} \tag{39}$$

The behavior of the wave functions $f(\rho), g(\rho)$ near the "event horizons" is structurally similar to the behavior of the radial functions in the case of the Schwarzschild and Reissner-Nordström fields [4], [5], [6].

For $\rho \to \rho_+$ $(\rho > \rho_+)$,

$$\begin{aligned} f &= A \sin(M_+ \ln(\rho - \rho_+) + \varphi_+) \\ g &= A \cos(M_+ \ln(\rho - \rho_+) + \varphi_+). \end{aligned} \tag{40}$$

In (40),

$$M_+ = \frac{\rho_+^2}{2\sqrt{\alpha^2 - \alpha_a^2}} \left( \varepsilon \left( 1 + \frac{\alpha_a^2}{\rho_+^2} \right) - \frac{m_\varphi \alpha_a}{\rho_+^2} \right). \tag{41}$$

For $\rho \to \rho_-$ $(\rho < \rho_-)$,

$$\begin{aligned} f &= -B \sin(M_- \ln(\rho_- - \rho) + \varphi_-) \\ g &= B \cos(M_- \ln(\rho_- - \rho) + \varphi_-). \end{aligned} \tag{42}$$

In (42),

$$M_- = \frac{\rho_-^2}{2\sqrt{\alpha^2 - \alpha_a^2}} \left( \varepsilon \left(1 + \frac{\alpha_a^2}{\rho_-^2}\right) - \frac{m_\varphi \alpha_a}{\rho_-^2} \right). \tag{43}$$

With $\rho \to 0$, the expressions for asymptotics are obtained in [16].

In (39), (40), (42) $C, A, B, \varphi_+, \varphi_-$ are the constant of integration.

Similarly to the case of the Schwarzschild and Reissner-Nordström fields [4] - [6], the oscillating functions $f$ and $g$ are ill defined at the external and internal "event horizons", but they are quadratically integrable functions at $\rho \neq \rho_+$ $(\rho > \rho_+)$ and at $\rho \neq \rho_-$ $(\rho < \rho_-)$.

## 2.4. Hermiticity of Chandrasekhar's Hamiltonian, boundary conditions for wave functions

According to the general theorem proven in [2], Chandrasekhar's stationary Hamiltonian is pseudo-Hermitian, or, in other words, Hermitian with Parker's weight operator (17).

In [4], [5], [6], Hermiticity of initial Hamiltonians for the Schwarzschild and Reissner-Nordström fields, considering the behavior of the wave functions, was established using an explicit form of spherical harmonics for spin-half particles. In our case with the angular functions $\overset{(-)}{S}(\theta), \overset{(+)}{S}(\theta)$ Hermiticity of the Hamiltonian in (11) $((\Phi, H\Psi) = (H\Phi, \Psi))$ can be also proven directly.

The boundary conditions for the wave functions are determined by the fulfillment of the Hilbert condition (7).

It follows from (7) that the defined combination of wave functions of the Hamiltonian (9) must be equal to zero on the surfaces of the inner or the outer ergospheres.

Because of the Hermiticity of the initial Hamiltonian and self-conjugacy of the Hamiltonian (9), the sought energy spectrum will be stationary and real.

## 2.5  Extreme Kerr field and naked singularity

The extreme field occurs if $\alpha = \alpha_a$ $\left(a = \frac{r_0}{2}\right)$. In this case, the external and internal "event horizons" coincide, their radius being equal to

$$\rho_+ = \rho_- = \alpha. \tag{44}$$

The radii of the outer and inner ergosphere surfaces equal

$$(r_{erg})_{1,2} = \frac{r_0}{2}(1 \pm \sin\theta). \tag{45}$$

For the case of interest, the system of equations (30) for the radial wave functions $f(\rho), g(\rho)$ and $f(\rho), g(\rho)$ reduces to

$$\left(1 - \frac{\alpha}{\rho}\right)^2 \frac{df}{d\rho} + \lambda \frac{\left|1 - \frac{\alpha}{\rho}\right|}{\rho} f - \left(\varepsilon\left(1 + \frac{a^2}{\rho^2}\right) - \frac{m_\varphi \alpha}{\rho^2} + \left|1 - \frac{\alpha}{\rho}\right|\right) g = 0$$

$$\left(1 - \frac{\alpha}{\rho}\right)^2 \frac{dg}{d\rho} - \lambda \frac{\left|1 - \frac{\alpha}{\rho}\right|}{\rho} g + \left(\varepsilon\left(1 + \frac{a^2}{\rho^2}\right) - \frac{m_\varphi \alpha}{\rho^2} - \left|1 - \frac{\alpha}{\rho}\right|\right) f = 0. \quad (46)$$

The wave functions are defined on the interval $\rho \in (0, \infty)$ except for the neighborhood of the "event horizon".

Let us consider an interesting case of a naked singularity that occurs if $\alpha_a > \alpha$ $\left(a > \frac{r_0}{2}\right)$.

In this case, the external and internal "event horizons" disappear, and the quantities $\rho_+, \rho_-$ in (42), (43) become complex numbers.

With increase in the angular momentum of rotation, $a > \frac{r_0}{2}$, the domain of $(\rho, \theta)$, where the condition $g_{00} > 0$ (7) is not fulfilled, reduces. In the limit of an infinitely high momentum $a$, the constraint on the domain of the wave function $\psi(r, \theta, \varphi)$ reduces to a solid of revolution about the z axis of a planar figure, which is infinitely narrow in the angle $\theta$ (Fig. 1d), near the equatorial plane with $0 \le r \le r_0$.

Because the Hilbert condition $g_{00} > 0$ constrains the domain of the wave function of the Dirac equation in the Kerr field also in the case of the naked singularity, the analysis of the energy levels of Dirac particles requires special consideration with numerical calculations of Eqs. (30), (37).

### 3. Kerr-Newman gravitational field

The Kerr-Newman solution of the general relativity equations is characterized by an electrically charged point source of gravitational field of mass $M$ and charge $Q$, which rotates with angular momentum $\mathbf{J} = Mc\mathbf{a}$.

## 3.1 Kerr-Newman solution in the Boyer-Lindquist coordinates

$$ds^2 = \left(1 - \frac{r_0 r - r_Q^2}{\rho_K^2}\right)dt^2 + \frac{2a\left(r_0 r - r_Q^2\right)}{\rho_K^2}\sin^2\theta \, dt \, d\varphi - \frac{\rho_K^2}{\Delta_{K-N}}dr^2 - \rho_K^2 d\theta^2 - $$
$$-\left(r^2 + a^2 + \frac{a^2\left(r_0 r - r_Q^2\right)}{\rho_K^2}\sin^2\theta\right)\sin^2\theta \, d\varphi^2. \tag{47}$$

Expression (47), as opposed to the expression for the Kerr solution (6), contains new notation related to the presence of electrical charge $Q$:

$$r_Q = \frac{\sqrt{G}Q}{c^2}; \quad \Delta_{K-N} = r^2 f_{K-N} = r^2\left(1 - \frac{r_0}{r} + \frac{a^2}{r^2} + \frac{r_Q^2}{r^2}\right).$$

In accordance with the Hilbert condition $(g_{00} > 0)$ in (47) it is meant that the following inequality holds:

$$\left(1 - \frac{r_0 r - r_Q^2}{\rho_K^2}\right) > 0. \tag{48}$$

The equality of the expression (48) to zero determines the outer and the inner surfaces of ergospheres of the Kerr-Newman field.

## 3.2 Separation of variables, equations and asymptotics for radial wave functions

Using Chandrasekhar's approach, Page [17] separated variables in the Dirac equation in the Kerr-Newman field with the solution (47) using the Penrose-Newman two-component spinor formalism [11] and the Kinnersley unit tetrad [12].

As a result, the equations for the angular functions $\overset{(-)}{S}(\theta), \overset{(+)}{S}(\theta)$ remain the same as for the Kerr field (see (23)). Equations for the radial functions $\overset{(-)}{R}_{K-N}(r), \overset{(+)}{R}_{K-N}(r)$ are written as

$$\Delta_{K-N}^{1/2}\left(\frac{d}{dr} - i\frac{K_{K-N}}{\Delta_{K-N}}\right)\overset{(-)}{R}_{K-N}(r) = (\lambda + imr)\overset{(+)}{R}_{K-N}(r),$$
$$\Delta_{K-N}^{1/2}\left(\frac{d}{dr} + i\frac{K_{K-N}}{\Delta_{K-N}}\right)\overset{(+)}{R}_{K-N}(r) = (\lambda - imr)\overset{(-)}{R}_{K-N}(r). \tag{49}$$

In Eqs. (49),

$$K_{K-N} = (r^2 + a^2)E - m_\varphi a + eQr.$$

It follows from Eqs. (49) that $\overset{(+)}{R^*}_{K-N}(r) = \overset{(-)}{R}_{K-N}(r)$ Then, for the real functions

$$g_1(r) = \overset{(-)}{R}_{K-N}(r) + \overset{(+)}{R}_{K-N}(r),$$
$$f_1(r) = -i\left(\overset{(-)}{R}_{K-N}(r) - \overset{(+)}{R}_{K-N}(r)\right). \tag{50}$$

Eqs. (49) take the following form:

$$f_{K-N}\frac{df_1}{dr} + \lambda\frac{\sqrt{f_{K-N}}}{r}f_1 - \left(E\left(1+\frac{a^2}{r^2}\right) - \frac{m_\varphi a}{r^2} + \frac{eQ}{r} + m\sqrt{f_{K-N}}\right)g_1 = 0,$$
$$f_{K-N}\frac{dg_1}{dr} + \lambda\frac{\sqrt{f_{K-N}}}{r}g_1 + \left(E\left(1+\frac{a^2}{r^2}\right) - \frac{m_\varphi a}{r^2} + \frac{eQ}{r} - m\sqrt{f_{K-N}}\right)f_1 = 0. \tag{51}$$

In Eqs. (51),

$$f_{K-N} = 1 - \frac{r_0}{r} + \frac{a^2 + r_Q^2}{r^2}. \tag{52}$$

The condition (48) leads to the necessity of considering only positive values of $f_{K-N} > 0$.

First, we consider the case

$$r_0 \geq 2\sqrt{a^2 + r_Q^2}. \tag{53}$$

In this case, the quantity $f_{K-N}$ can be represented as

$$f_{K-N} = \left(1 - \frac{r^+_{K-N}}{r}\right)\left(1 - \frac{r^-_{K-N}}{r}\right), \tag{54}$$

where

$$r^\pm_{K-N} = \frac{r_0}{2} \pm \sqrt{\frac{r_0^2}{4} - a^2 - r_Q^2}. \tag{55}$$

The quantities $r^\pm_{K-N}$ are outer and inner radii of the "event horizons" of the Kerr-Newman field.

In the dimensionless variables $\varepsilon = \frac{E}{m}$; $\rho = \frac{r}{l_c}$; $2\alpha = \frac{r_0}{l_c} = \frac{2GMm}{\hbar c}$; $\alpha_Q = \frac{r_Q}{l_c} = \frac{\sqrt{G}Qm}{\hbar c}$;

$\alpha_{em} = \frac{eQ}{\hbar c}$, the expressions (54), (55) can be represented as

$$f_{K-N} = \left(1 - \frac{\rho^+_{K-N}}{\rho}\right)\left(1 - \frac{\rho^-_{K-N}}{\rho}\right), \tag{56}$$

$$\rho^\pm_{K-N} = \alpha \pm \sqrt{\alpha^2 - \alpha_a^2 - \alpha_Q^2}. \tag{57}$$

Eqs. (51) in the dimensionless variables have the following form:

$$\left(1-\frac{\rho_{K-N}^+}{\rho}\right)\left(1-\frac{\rho_{K-N}^-}{\rho}\right)\frac{df_1}{d\rho} + \lambda \frac{\sqrt{\left(1-\frac{\rho_{K-N}^+}{\rho}\right)\left(1-\frac{\rho_{K-N}^-}{\rho}\right)}}{\rho} f_1 -$$

$$-\left(\varepsilon\left(1+\frac{\alpha_a^2}{\rho^2}\right) - \frac{m_\varphi \alpha_a}{\rho^2} + \frac{\alpha_{em}}{\rho} + \sqrt{\left(1-\frac{\rho_{K-N}^+}{\rho}\right)\left(1-\frac{\rho_{K-N}^-}{\rho}\right)}\right) g_1 = 0,$$

(58)

$$\left(1-\frac{\rho_{K-N}^+}{\rho}\right)\left(1-\frac{\rho_{K-N}^-}{\rho}\right)\frac{dg_1}{d\rho} - \lambda \frac{\sqrt{\left(1-\frac{\rho_{K-N}^+}{\rho}\right)\left(1-\frac{\rho_{K-N}^-}{\rho}\right)}}{\rho} g_1 +$$

$$+\left(\varepsilon\left(1+\frac{\alpha_a^2}{\rho^2}\right) - \frac{m_\varphi \alpha_a}{\rho^2} + \frac{\alpha_{em}}{\rho} - \sqrt{\left(1-\frac{\rho_{K-N}^+}{\rho}\right)\left(1-\frac{\rho_{K-N}^-}{\rho}\right)}\right) f_1 = 0.$$

Similarly to the Kerr field, in accordance with the condition (48), the domain for the wave functions $f_1(\rho), g_1(\rho)$ in Eqs. (58) is the space $(\rho, \theta, \varphi)$, the radii of which are larger than the radii of the outer ergosphere surface and the radii $\rho(\theta)$ are smaller than the radii of the inner ergosphere surface of the Kerr-Newman field.

The asymptotic behavior of the radial functions for the finite motion is structurally the same as for the Kerr field.

For $\rho \to \infty$ leading terms of asymptotics equals

$$f_1 = Ce^{-\rho\sqrt{1-\varepsilon^2}},$$
$$g_1 = -\sqrt{\frac{1-\varepsilon}{1+\varepsilon}} f_1.$$

(59)

For $\rho \to \rho_{K-N}^+ \ (\rho > \rho_{K-N}^+)$,

$$f_1 = A\sin\left(M_{K-N}^+ \ln\left(\rho - \rho_{K-N}^+\right) + \varphi_{K-N}^+\right),$$
$$g_1 = A\cos\left(M_{K-N}^+ \ln\left(\rho - \rho_{K-N}^+\right) + \varphi_{K-N}^+\right),$$

(60)

where

$$M_{K-N}^+ = \frac{\left(\rho_{K-N}^+\right)^2}{2\sqrt{\alpha^2 - \alpha_a^2 - \alpha_Q^2}}\left(\varepsilon\left(1+\frac{\alpha_a^2}{\left(\rho_{K-N}^+\right)^2}\right) - \frac{m_\varphi}{\left(\rho_{K-N}^+\right)^2} + \frac{\alpha_{em}}{\rho_{K-N}^+}\right).$$

(61)

For $\rho \to \rho_{K-N}^- \ (\rho < \rho_{K-N}^-)$,

$$f_1 = -B\sin\left(M_{K-N}^- \ln\left(\rho_{K-N}^- - \rho\right) + \varphi_{K-N}^-\right),$$
$$g_1 = B\cos\left(M_{K-N}^- \ln\left(\rho_{K-N}^- - \rho\right) + \varphi_{K-N}^-\right),$$

(62)

where

$$M_{K-N}^{-} = \frac{\left(\rho_{K-N}^{-}\right)^2}{2\sqrt{\alpha^2 - \alpha_a^2 - \alpha_Q^2}} \left( \varepsilon\left(1 + \frac{\alpha_a^2}{\left(\rho_{K-N}^{-}\right)^2}\right) - \frac{m_\varphi}{\left(\rho_{K-N}^{-}\right)^2} + \frac{\alpha_{em}}{\rho_{K-N}^{-}} \right). \quad (63)$$

With $\rho \to 0$, the expressions for asymptotics are obtained in [16].

In (59), (60), (62) $C, A, B, \varphi_{K-N}^{+}, \varphi_{K-N}^{-}$ are the constants of integration.

Similarly to the case of the Schwarzschild [4], [5], Reissner-Nordströmf [6] and Kerr field (see (40) - (43)), the oscillating functions $f_1$ and $g_1$ are ill defined at the external and internal "event horizons", but they are quadratically integrable functions at $\rho \neq \rho_{K-N}^{+}$ $\left(\rho > \rho_{K-N}^{+}\right)$ and at $\rho \neq \rho_{K-N}^{-}$ $\left(\rho < \rho_{K-N}^{-}\right)$.

Hermiticity of the Dirac Hamiltonian in the Kerr-Newman field is proven similarly to the Kerr field ((Sect. 2.4) above, see also [2], [18]).

The boundary conditions for the wave functions are determined similarly to the case of the Kerr field (Sect. 2.4).

As a result, we can conclude that the system of equations (58) has a stationary real energy spectrum. The spectrum and the wave functions will be determined later in numerical calculations.

### 3.3 Extreme Kerr-Newman field and naked singularity

The extreme field occurs if $\alpha = \sqrt{\alpha_a^2 + \alpha_Q^2}$ $\left(\frac{r_0}{2} = \sqrt{a^2 + r_Q^2}\right)$. In this case, the external and internal "event horizons" coincide, their radius being equal to

$$\rho_{K-N}^{+} = \rho_{K-N}^{-} = \alpha. \quad (64)$$

The radii of the outer and inner ergospheres equal

$$\left(r_{erg}\right)_{1,2} = \frac{r_0}{2}\left(1 \pm \frac{2a}{r_0}\sin\theta\right). \quad (65)$$

For the case of interest, the system of equations (58) reduces to

$$\left(1 - \frac{\alpha}{\rho}\right)^2 \frac{df_1}{d\rho} + \lambda \frac{\left|1 - \frac{\alpha}{\rho}\right|}{\rho} f_1 - \left(\varepsilon\left(1 + \frac{\alpha_a^2}{\rho^2}\right) - \frac{m_\varphi \alpha_a}{\rho^2} + \frac{\alpha_{em}}{\rho} + \left|1 - \frac{\alpha}{\rho}\right|\right) g_1 = 0,$$

$$\left(1 - \frac{\alpha}{\rho}\right)^2 \frac{dg_1}{d\rho} - \lambda \frac{\left|1 - \frac{\alpha}{\rho}\right|}{\rho} g_1 + \left(\varepsilon\left(1 + \frac{\alpha_a^2}{\rho^2}\right) - \frac{m_\varphi \alpha_a}{\rho^2} + \frac{\alpha_{em}}{\rho} - \left|1 - \frac{\alpha}{\rho}\right|\right) f_1 = 0. \quad (66)$$

The wave functions are defined on the interval $\rho \in (0, \infty)$ except for the neighborhood of the "event horizon".

The case of naked singularity occurs if $\alpha < \sqrt{\alpha_a^2 + \alpha_Q^2}$ $\left(\dfrac{r_0}{2} < \sqrt{a^2 + r_Q^2}\right)$.

In this case, the external and internal "event horizons" disappear, and the quantities $\rho_{K-N}^+$ and $\rho_{K-N}^-$ become complex numbers.

As noted in Sect. 2.5 above, the analysis of the energy levels of Dirac particles in the case of the naked singularity of the Kerr-Newman field requires special consideration with numerical calculations of Eqs. (51), (58) with corresponding boundary conditions.

Figs. 1, 2 in coordinates $\left(r' = \dfrac{2r}{r_0}, \theta\right)$ show the domains of the wave functions of the Dirac equation in the Kerr and Kerr-Newman fields for some values of $a' = \dfrac{2a}{r_0}$ and $r_Q' = \dfrac{2r_Q}{r_0}$.

The color areas correspond to the domains, where the wave functions should be zero to ensure the fulfillment of the Hilbert condition $g_{00} > 0$.

The surfaces of the outer and inner ergospheres are in color heavy lines.

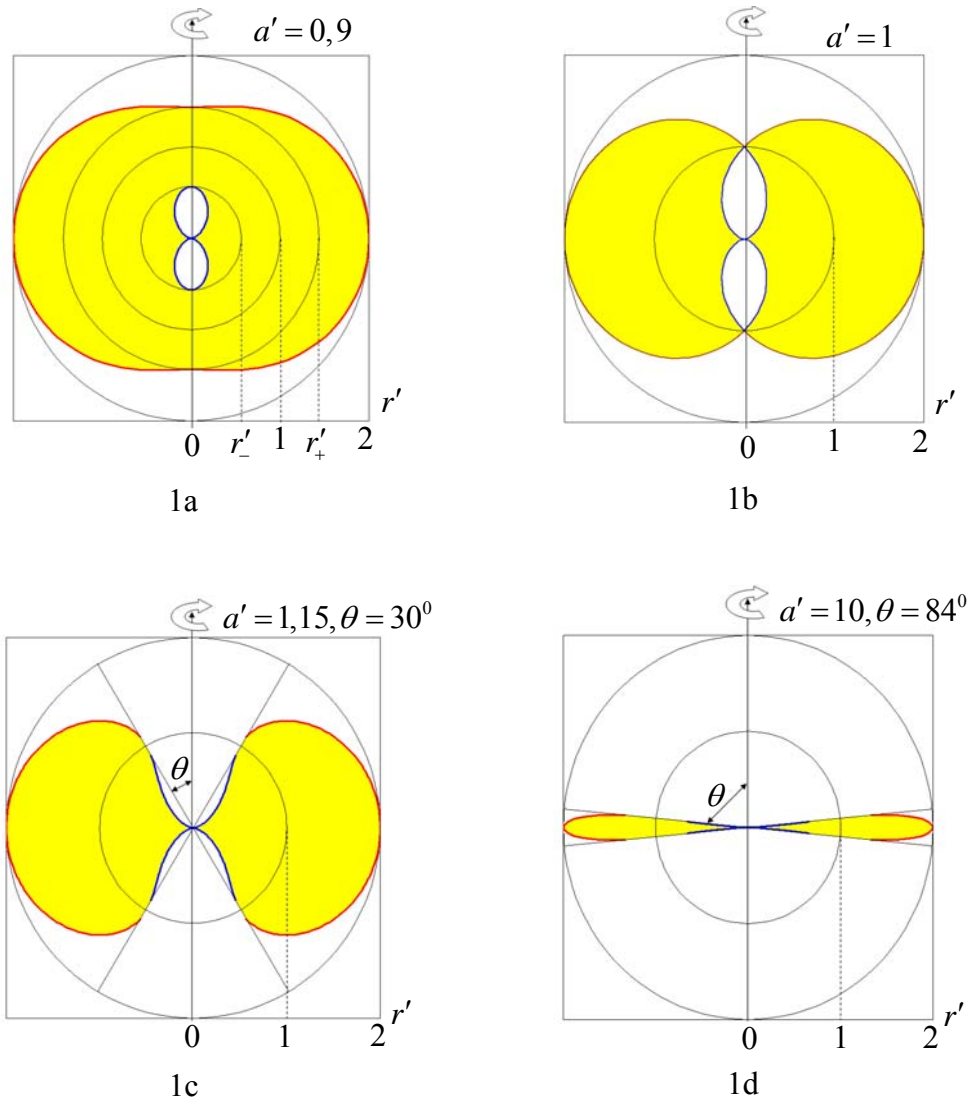

Fig. 1. The domains of the wave functions of the Dirac equation in the Kerr field for some values of $a' = \dfrac{2a}{r_0}$.

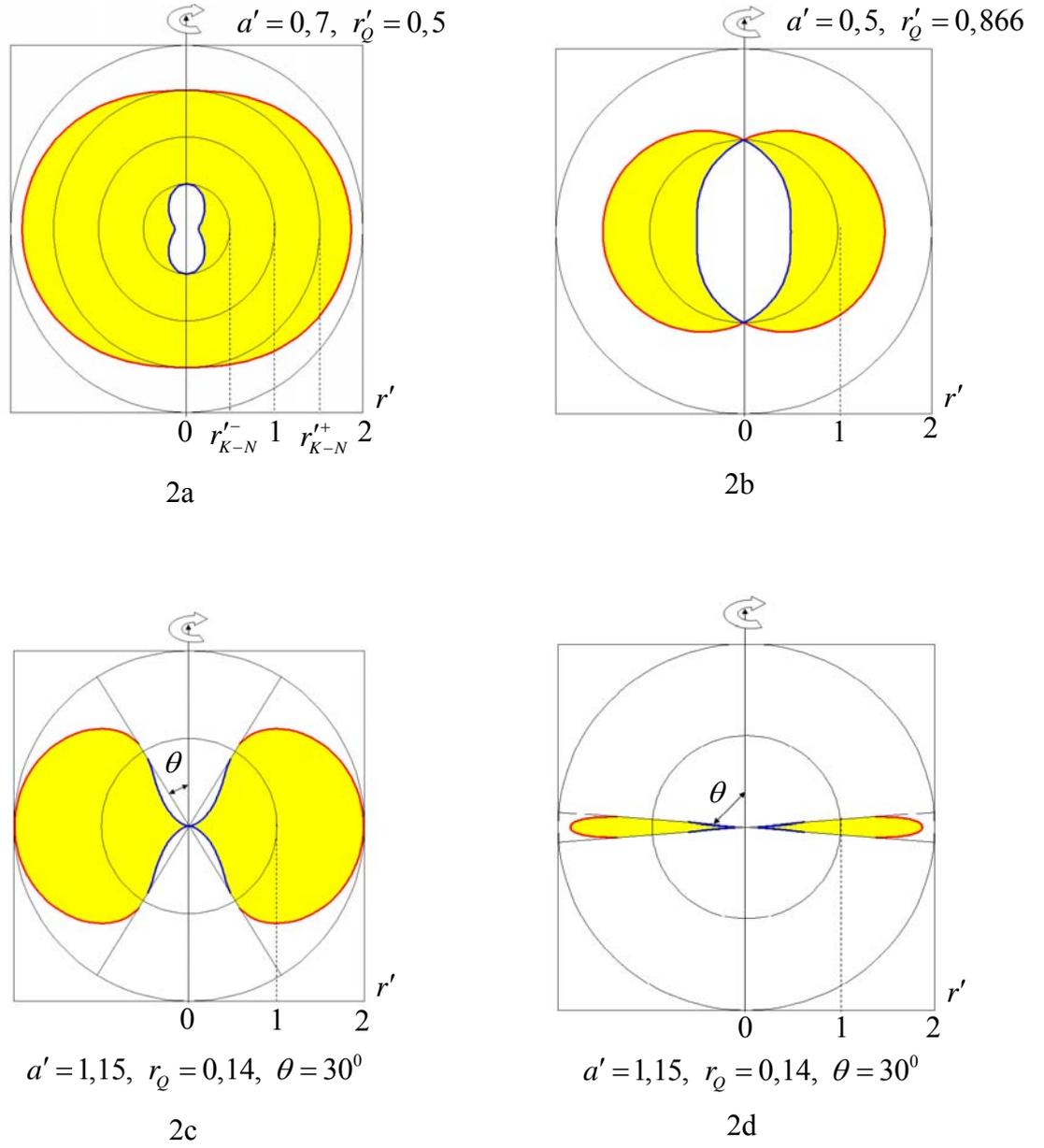

Fig. 2. The domains of the wave functions in the Kerr-Newman field for some values of

$$a' = \frac{2a}{r_0},\ a'_Q = \frac{2r_Q}{r_0}.$$

## 4. Conclusions

Based on the results of this work we can draw the following conclusions:
1. We have proved the possibility of existence of stationary bound states of spin-half particles for the Kerr and Kerr-Newman solutions.
2. Bound states with a real energy spectrum can exist both above the surfaces of the outer ergospheres and under the surfaces of the inner ergospheres of the Kerr and Kerr-Newman fields.
3. Provided that the Hilbert condition ($g_{00} > 0$) is satisfied, the surfaces of the outer and inner ergospheres play the role of infinitely high potential barriers, not allowing quantum-mechanical Dirac particles to cross them.
4. The wave function of spin-half particles in the range between the surfaces of the outer and inner ergospheres is zero.

It follows from this that the existence of rotating collapsars of the new type is possible. These collapsars are:

- inert (Dirac particles cannot cross the surfaces of the outer and inner ergospheres);

- have no Hawking radiation property [7] (Hawking radiation requires that there is a wave function (Dirac field operators) between the surfaces of the inner and outer ergospheres [19] - [26]);

- at least with $\alpha \geq \sqrt{\alpha_a^2 + \alpha_Q^2}$ $\left(\dfrac{r_0}{2} \geq \sqrt{a^2 + r_Q^2}\right)$ they provide for the existence of stationary bound states of Dirac particles above the outer and under the inner ergosphere surfaces of the Kerr and Kerr-Newman fields.

Thus, the results of this study and works [4] - [6] can be useful for the improvement of some aspects of the standard cosmological model related to the evolution of the universe and interaction of rotating collapsars with surrounding matter.

### Acknowledgement

We thank A.L. Novoselova, Yu.V.Petrov for the substantial technical help in the preparation of this paper.